\documentclass[10pt,prl,aps,twocolumn,showpacs,floatfix]{revtex4}
\usepackage{graphicx}
\usepackage{amssymb}
\usepackage{amsmath}

\begin{document}

\title{Confinement and Deconfinement of Spinons in Two Dimensions}

\author{Ying Tang and Anders W. Sandvik}
\affiliation{Department of Physics, Boston University, 590 Commonwealth Avenue, Boston, Massachusetts 02215}

\begin{abstract}
We use Monte Carlo methods to study spinons in two-dimensional quantum spin systems, characterizing their intrinsic size $\lambda$ 
and confinement length $\Lambda$. We confirm that spinons are deconfined, $\Lambda \to \infty$ and $\lambda$ finite, in a resonating 
valence-bond spin-liquid state. In a valence-bond solid, we find finite $\lambda$ and $\Lambda$, with $\lambda$ of a single spinon 
significantly larger than the bound-state---the spinon is soft and shrinks as the bound state is formed. Both $\lambda$ and $\Lambda$ 
diverge upon approaching the critical point separating valence-bond solid and N\'eel ground states. We conclude that the spinon deconfinement is 
marginal in the lowest-energy state in the spin-$1$ sector, due to weak attractive spinon interactions. Deconfinement in the vicinity 
of the critical point should occur at higher energies.
\end{abstract}

\date{\today}


\maketitle

The concept of {\it confinement} originated in particle physics, where quarks are bound into hadrons by ``gluon strings'', leading to a distance-independent 
force and the inability to isolate individual quarks \cite{Munzinger}. This picture has been carried over to condensed matter physics as well. One example
is the non-magnetic valence-bond-solid (VBS) state of a two-dimensional (2D) quantum spin system, whose ground state consists of a crystaline arrangement 
of two-spin singlets (valence bonds). An elementary excitation corresponds to promoting one bond into a triplet. The members of this triplet (the spinons) 
stay bound to each other, due to a string of misaligned singlets forming when separating the spinons \cite{Read}. In a one-dimensional (1D) VBS the confining 
string does not exist, as there is no medium surrounding the phase-shifted dimers between the spinons. The spinons then deconfine and become independent 
elementary excitations. The spinon concept was originally introduced as a spin analog \cite{Shastry} of the fractional excitations of conjugated polymers 
\cite{Su}, which are phase-twist solitons---domain walls between two out-of-phase dimerized chain segment. 

An ongoing quest in quantum magnetism is to identify 2D systems in which spinons can deconfine. Deconfinement and weak confinement 
of spinons have been experimentally observed in one dimensional (1D) spin chains \cite{Tennant} and ladders \cite{Lake}, respectively, by neutron scattering. 
Deconfinement is characterized by a broad continuum of spin-$1$ excitations. These observations and conclusions are aided by the known triplet spectrum 
\cite{Pereira} of the Heisenberg chain, where the existence of spinons is understood rigorously \cite{Faddeev}. For 2D systems, this way of detection is 
difficult, due to a lack of established model spectral functions for spinons and the existence of significant continua even in the magnon spectrum of 
antiferromagnets \cite{Sandvik4}. While alternative experimental methods have also been proposed \cite{Zhou}, this remains a challenging issue. In numerical 
model studies, correlations of excess magnetization with impurities \cite{Poilblanc,Doretto} and other inhomogeneities \cite{Yan} have been used to study spinons. 

Here we use a quantum Monte Carlo (QMC) approach for homogeneous systems \cite{Tang1} and address the mechanism of 2D spinon deconfinement 
promoted by Senthil {\it et al.} \cite{Senthil1,Senthil2,Levin}. In their picture, a spinon in a square-lattice VBS can be viewed as vortex at the nexus of four 
VBS domains, with the core carrying spin-$1/2$. These vortices bind with anti-vortices to form the gapped spin-$1$ triplons. The binding potential weakens 
as the fluctuations of the VBS increase and a {\it deconfined quantum-critical} (DQC) point is approached which separates the VBS from the standard 
antiferromagnetic (N\'eel) state. 

Deconfined spinons are expected in spin liquids, which are non-magnetic states with no broken symmetries \cite{Sachdev}. The DQC point is an 
isolated, gapless (algebraic) spin liquid. In the near-DQC VBS, the triplon is predicted to become large, exceeding the correlation length \cite{Senthil2,Levin}, 
and may be regarded as a pair of loosely bound spinons. Several QMC studies of $J$-$Q$ models, where the Heisenberg exchange is supplemented with 
higher-order interactions \cite{Sandvik1}, support the existence of an unusual critical point \cite{Sandvik1,Kaul,Lou1,Sandvik2,Banerjee} and the critical 
thermodynamic properties are also consistent with spinons \cite{Sandvik3}. 

{\it Objectives and Findings}---We here study properties of the DQC spinons and their near-DQC bound states in the 2D $J$-$Q$ model. Computing the spinon 
size and confinement length, we demonstrate that a naive picture of a large bound state of two small objects fails---the spinons themselves 
are of size comparable to the bound state and are ``soft'', shrinking when forming bound states. This differs from the prototypical short-bond 
resonating valence-bond (RVB) spin liquid \cite{Anderson1,Sutherland}, which we also study as a point of reference. Here the 
deconfined spinons are small, with radius of a few lattice spacings.

\begin{figure}
\centerline{\includegraphics[width=4.2cm, clip]{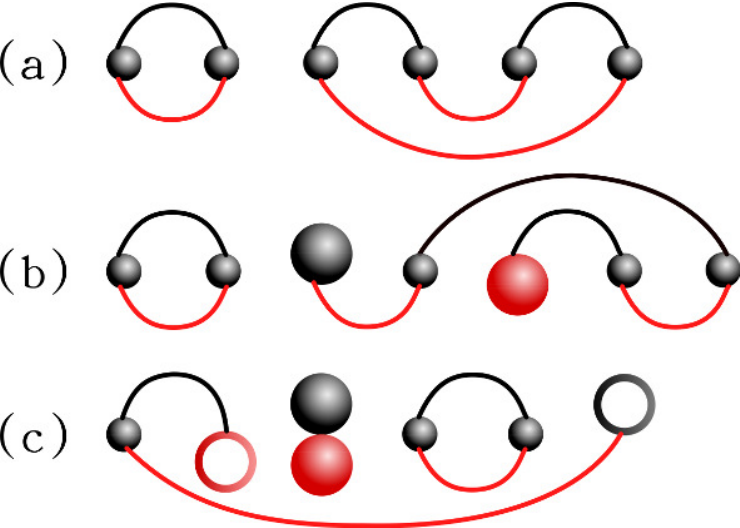}}
\vskip-2mm
\caption{(Color online) Transition graphs for $S=0$ (a), $S=1/2$ (b), and $S=1$ (c). The bonds and larger circles indicate valence bonds 
and unpaired ($\uparrow$) spins, respectively, in ket (darker, black) and bra (lighter, red) states. The statistics of separation of unpaired spins in 
(b) and (c) can be used to characterize the spatial properties of spinons and their bound states.}
\label{fig1}
\vskip-3mm
\end{figure}

{\it Methods---}In Ref.~\cite{Tang1} we showed how spinons can be characterized for bipartite systems in the valence-bond (singlet) 
basis extended to $S=1/2$ and $S=1$ by introducing one (on a lattice with an odd number of sites) or two (for an even number of sites) unpaired spins. 
An $S=0$ state is a superposition of states $|V_N\rangle$ in which all $N$ spins are paired up into $N/2$ singlets between sublattice $A$ and $B$ sites, 
while for $S=1/2$ and $S=1$ the basis states are, respectively, $|V_{N-1}\rangle \otimes |\uparrow_a\rangle$ and 
$|V_{N-2}\rangle \otimes |\uparrow_a\rangle \otimes |\uparrow_b\rangle$, where $a$ and $b$ are sites on different sublattices \cite{Wang,Banerjee2}. 
Superpositions can be generated probabilistically either in variational states (amplitude-product states \cite{Liang} and their generalizations \cite{Lin}) or 
in QMC simulations of Hamiltonians. Examples of configurations are shown in Fig.~\ref{fig1}. We use the formulation of the projector QMC method discussed 
in Ref.~\cite{Sandvik5}. 

The valence-bond basis being overcomplete and non-orthogonal, one should not consider the unpaired spins directly as spinons. However, 
as shown in Ref.~\cite{Tang1}, to which we refer for details, the statistics of the separation of unpaired spins in transition graphs (see Fig.~\ref{fig1}) are
related to wave-function overlaps and can be used to extract the intrinsic size of a single spinon as well as the size of an $S=1$ bound state. We denote 
the probability distribution between $\uparrow$ spins in the bra and ket state of an $S=1/2$ configuration $P_{AA}(r)$ and between those on different sublattices 
in an $S=1$ state by $P_{AB}(r)$. The former contains information on the spinon size and the latter on the bound state (or lack thereof). The same information 
is contained also in the difference $C(r)=C_S(r)-C_0(r)$ between the $z$-component spin-spin correlation function in the $S=1/2,1$ and $S=0$ states 
\cite{Tang1}. The method of statistics of separations is easier, since $C(r)$ is statistically noisy for large systems. The latter method 
can, however, be extended to non-bipartite systems and other computational methods.

{\it Models---}We use Monte Carlo sampling \cite{Liang} to generate valence-bond configurations for the short-bond RVB, as done in two recent studies to characterize 
its spin and dimer correlations \cite{Tang2,Alet}. We use the RVB as a reference point for an established U($1$) spin liquid. It is expected to have deconfined 
spinons but their spatial characteristics have not been considered before, to our knowledge. Our main interest is to characterize spinons at the N\'eel--VBS
transition of the $J$-$Q$ model. Here we consider a $Q$-term consisting of three singlet projectors $C_{ij}=1/4 - {\bf S_{i}} \cdot {\bf S_{j}}$. The Hamiltonian 
is \cite{Lou1}
\begin{equation}
H = -J\sum_{\langle ij\rangle}C_{ij} - Q_3\hskip-1.5mm\sum_{\langle ijklmn\rangle}\hskip-2.5mm C_{ij}C_{kl}C_{mn},
\label{JQ3}
\end{equation}
where $\langle ij\rangle$ and $\langle ijklmn\rangle$ denote, respectively, nearest neighbors and three nearest-neighbor pairs in horizontal or vertical columns
on the square lattice. The critical coupling ratio separating the N\'eel and VBS phases is $g_c=(Q_3/J)_c\approx 1.50$ \cite{Lou1}, and for large $g$ the 
VBS order is very robust. 

We also study the 1D $J$-$Q_3$ model including static dimerization, with different couplings on even and odd bonds;
\begin{eqnarray}
&&H=-\sum_{i={\rm even}} (J_1C_{2i,2i+1} + J_2 C_{2i+1,2i+2})~~~~ \nonumber \\
&&~~~~ -Q_3\sum_{i}C_{i,i+1}C_{i+1,i+2}C_{i+2,i+3}.
\label{j1j2}
\end{eqnarray}
When $J_1=J_2$ the quasi-ordered to VBS transition  (which is of the same kind as in the frustrated Heisenberg chain \cite{Sanyal}) is at $Q_3/J=0.1645$ \cite{Tang1}.
In this study we use this critical ratio and tune $\rho=J_2/J_1$. The explicit dimerization introduced by $\rho \neq 1$ leads to confinement \cite{Doretto}, while 
the spinons are deconfined when $\rho = 1$. We will demonstrate similarities with the 2D $J$-$Q_3$ model.

\begin{figure}
\centerline{\includegraphics[width=7.3cm, clip]{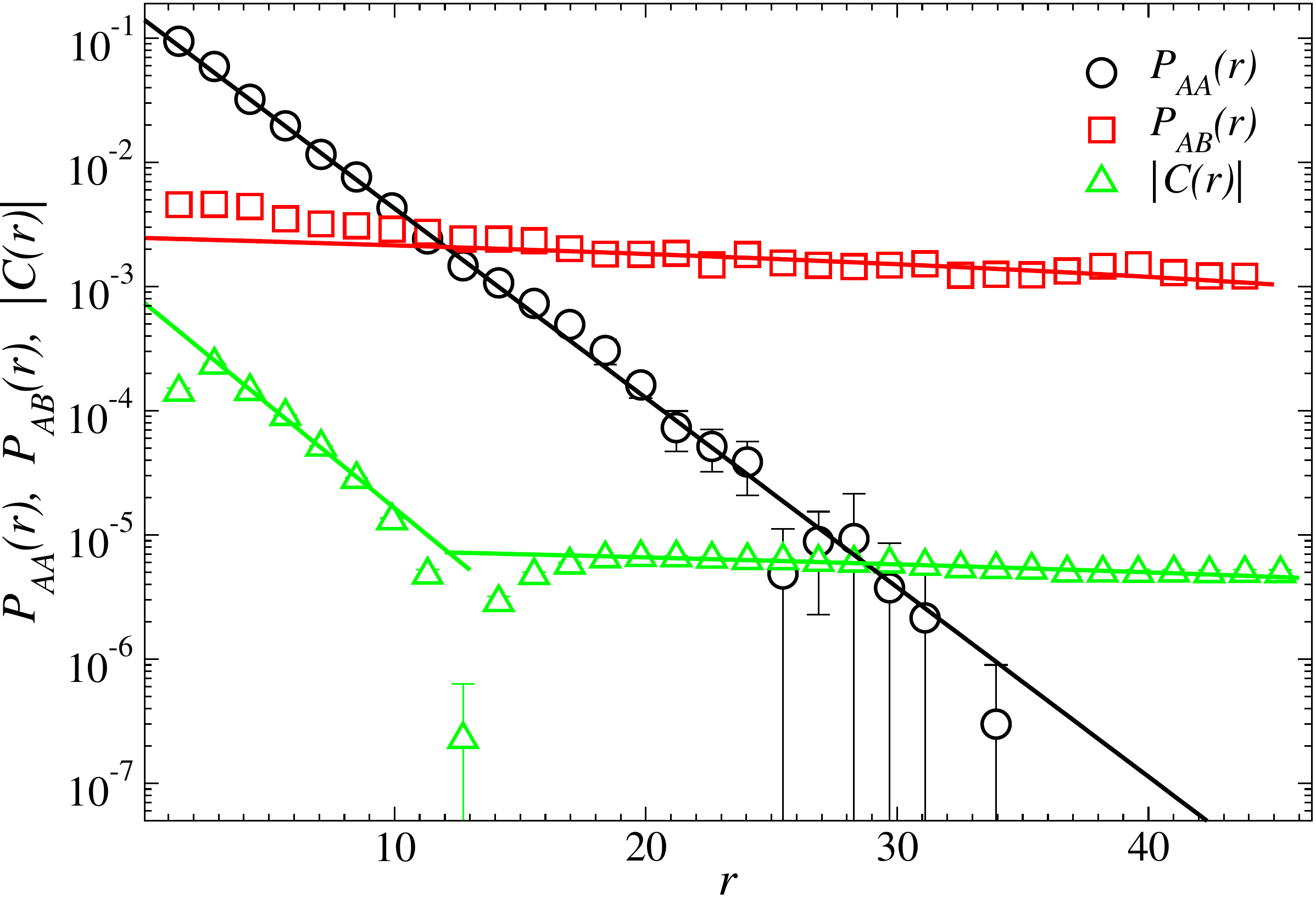}}
\vskip-2mm
\caption{(Color online) Spinon distributions and correlations along the diagonal lattice direction of the RVB spin liquid. Black circles 
show the single-spinon overlap $P_{AA}(r)$ in an $S=1/2$ state on a $65 \times 65$ lattice. The exponential decay (fitted line) gives the spinon size 
$\lambda=2.96(1)$ lattice spacings. Red squares show the two-spin distribution $P_{AB}({\bf r})$ in the $S=1$ state on a $64\times 64$ lattice, fitted
to $\sim 1/r^\alpha$ with $\alpha \approx 0.6$. The green triangles show the absolute value of the difference $C(r)$  between the spin correlations in the $S=1$ and $S=0$ systems. 
This quantity exhibits both a spinon-size effect (exponential short-distance decay) and deconfinement (weak power-law decay at long distances). There is
a phase shift at $r\approx9\sqrt{2}$.}
\label{fig2}
\vskip-1mm
\end{figure}

{\it RVB Spin Liquid}---RVB spin liquids have been considered as promising candidates for explaining high temperature superconductivity in cuprates when 
doped \cite{Anderson1,Anderson2}. It is therefore also interesting to examine in detail properties of the insulating host system. Recently, it was found that the 
simplest equal-amplitude short-bond RVB is a quantum-critical VBS, with exponentially decaying spin correlations but power-law dimer correlations \cite{Tang2,Alet}. 
It has been long expected that the RVB hosts deconfined spinon excitations \cite{Sachdev}. 

The parent Hamiltonian of the short-bond RVB was found recently \cite{Cano}. Although the $S=1/2$ and $S=1$ states we study here  may not be its exact lowest states 
in these sectors, one can still expect them to be good variational states---the actual excitations of the Cano-Fendley Hamiltonian should be very similar. 

We characterize the RVB spinons in Fig.~\ref{fig2}. The $S=1/2$ distribution $P_{AA}(r)$ demonstrates a well-defined intrinsic spinon 
wave packet, decaying as $e^{-r/\lambda}$ with the spinon size $\lambda = 2.96(1)$. The $S=1$ distribution $P_{AB}(r)$ is peaked at short distances and appears to decay 
as $r^{-\alpha}$ with $\alpha \approx 0.6$. This implies marginal deconfinement due to weak attractive spinon-spinon interactions.
In Fig.~\ref{fig2} we also show that the length-scales observed in $P_{AA}(r)$ and $P_{AB}(r)$ are manifested in the $S=0,1$ correlation function $C(r)$ as well.
Hence, the distributions do capture actual physical, basis-independent length-scales \cite{Tang1}.

\begin{figure}
\centerline{\includegraphics[width=7.3cm, clip]{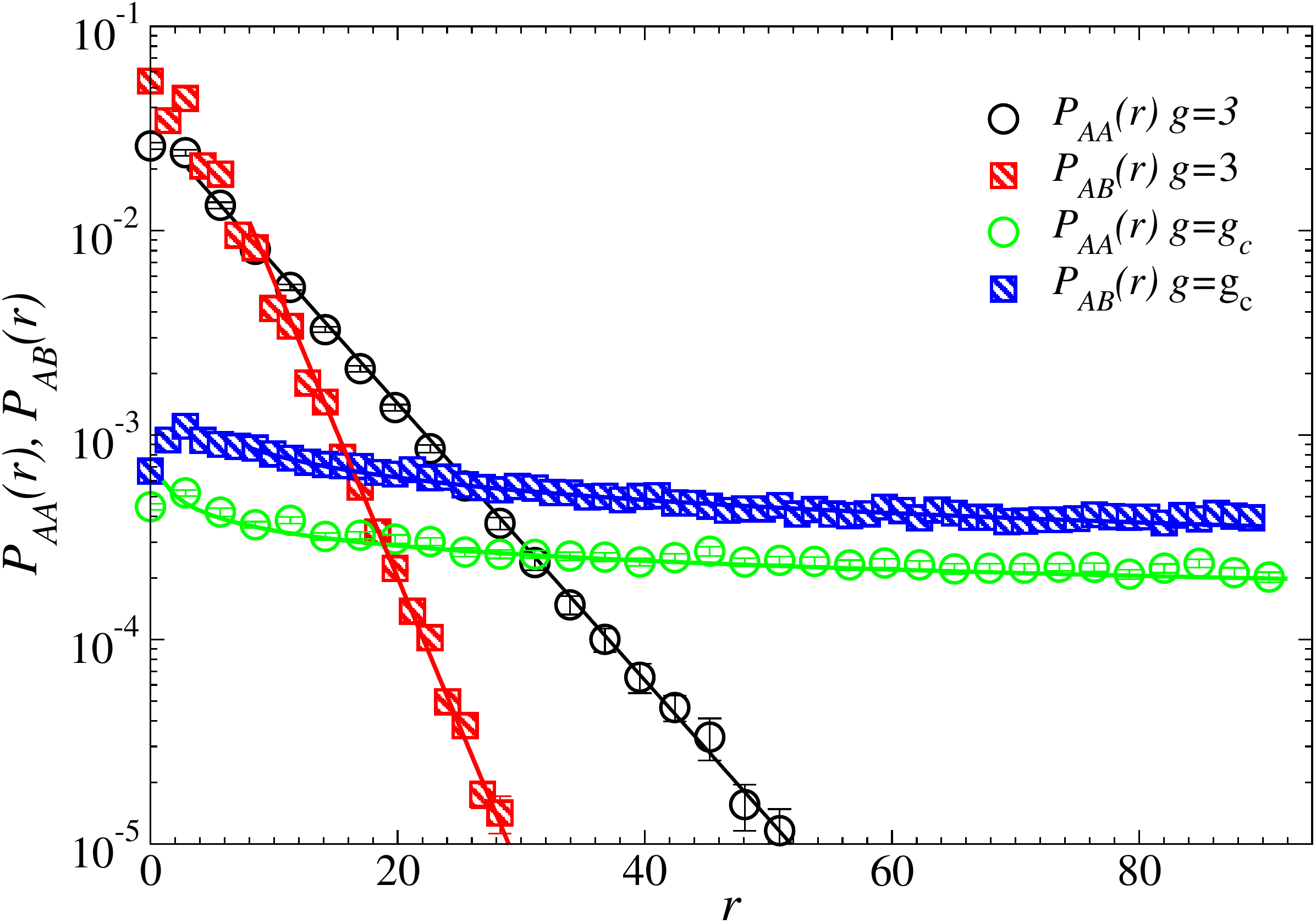}}
\vskip-2mm
\caption{(Color online) Spinon distributions for the 2D $J$-$Q_3$ model on $L=128$ and $129$ lattices at $g=3$ (VBS) and $g_c=1.5$ (critical).
The lines are exponential fits for the $g=3$ VBS, giving the single-spinon size $\lambda=6.4$ extracted from $P_{AA}(r)$ in the $S=1/2$ state and
the confinement length $\Lambda=3.1$ extracted from $P_{AB}(r)$ in the $S=1$ state. Both lengths diverge as $g \to g_c$, where the distributions decay
algebraically.}
\label{fig3}
\vskip-3mm
\end{figure}

{\it $J$-$Q$ model---}The N\'eel--VBS transition has been debated for years \cite{Read}. In 2004, Senthil et al.~presented a scenario encompassing
several earlier works \cite{Read,Murthy,Motrunich} and further proposed a mechanism leading to a generic continuous transition \cite{Senthil1, Senthil2, Levin}. 
This scenario is at odds with the ``Landau rule'' according to which transitions  between the two ordered states breaking unrelated symmetries should be first-order. A key aspect of 
the theory is that both order parameters arise out of spinons, which condense in the N\'eel state and confine in the VBS (where valence bonds 
can be regarded as tightly bound spinon pairs). Exactly at the DQC point separating the ordered states the spinons should deconfine. Although opposing views 
have been put forward \cite{Jiang,Kuklov1,Kuklov2}, the generic continuous nature of the N\'eel--VBS transition has support in QMC studies of $J$-$Q$ 
\cite{Sandvik1,Sandvik2,Lou1,Sandvik5} and other \cite{Kaul} models, including a predicted emergent U($1$) symmetry reflecting the gauge structure 
of the theory, where spinons interact with an U($1$) gauge field in a non-compact CP$^{1}$ action.

The DQC scenario motivates us to investigate spinons directly, by computing the spinon size $\lambda$ and the confinement length $\Lambda$ in VBS states and approaching 
the critical point.  As shown in Fig.~\ref{fig3}, both $P_{AA}(r)$ and $P_{AB}(r)$ are exponentially decaying in the VBS phase, with $\lambda=6.4(1)$ and $\Lambda=3.1(1)$ 
at $g=3$. Surprisingly, the intrinsic size of a single spinon is, thus, much larger that the bound state of two spinons.  We interpret this as a softness of the 
spinon, which causes it to shrink when subject to attractive interactions from an anti-spinon. This should be a signature of the vortex-nature of the spinon, 
as the opposite circulations of the members of the pair should lead cancellations away from the double-vortex core. Such shrinkage of vortices could in principle 
also occur under certain conditions in superconductors \cite{Chaves1,Chaves2,Babaev}.

\begin{figure}
\centerline{\includegraphics[width=7.3cm, clip]{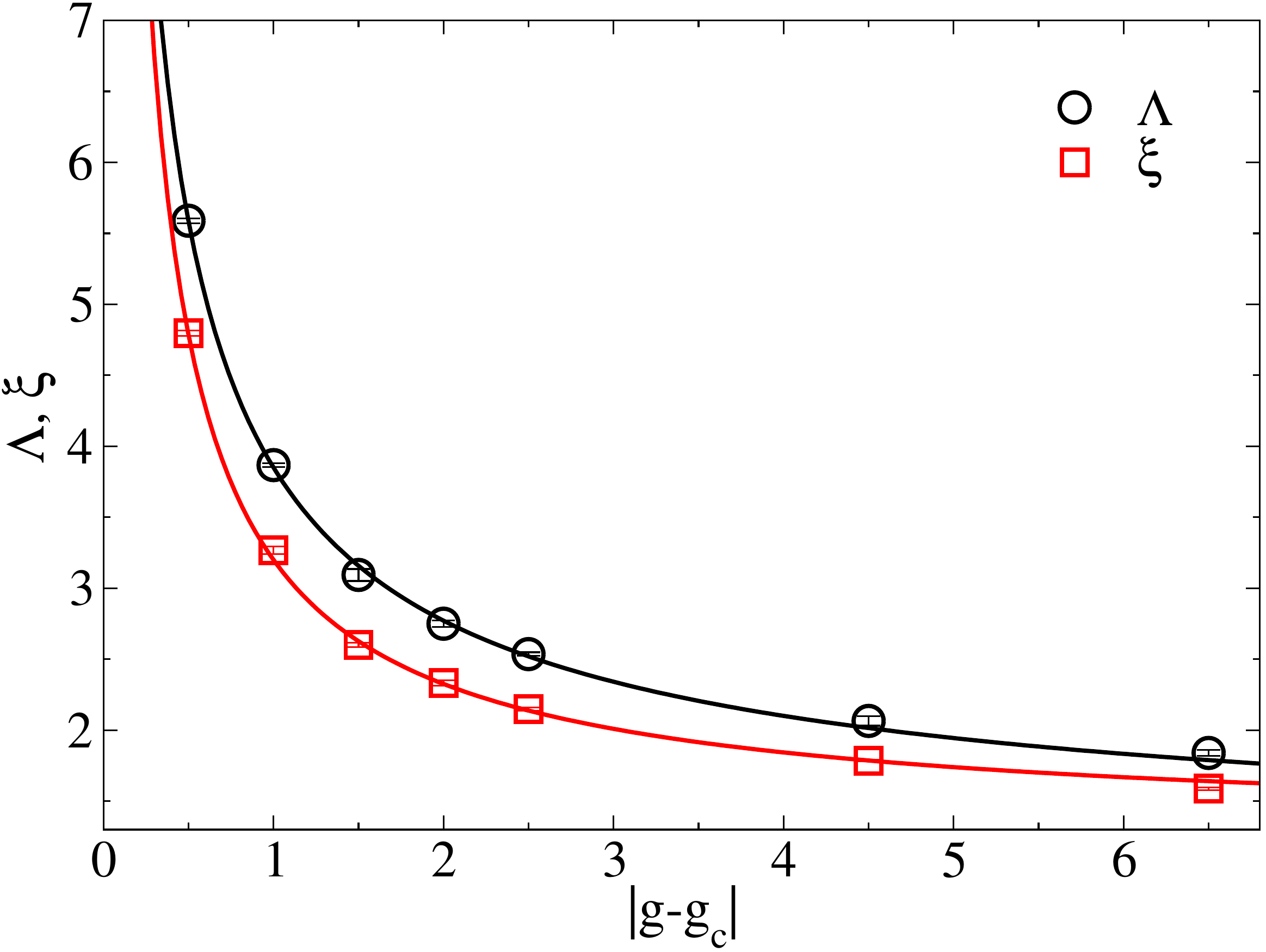}}
\vskip-2mm
\caption{(Color online) The confinement length $\Lambda$, the correlation length $\xi$, and the spinons size $\lambda$ in the VBS phase and approaching 
the critical point ($g_c=1.50$) of the $J$-$Q_3$ model. The calculations were done with $L=128$ and $129$. Thhe power-law fits (solid curves) are discussed 
in the text.}
\label{fig4}
\vskip-3mm
\end{figure}

Approaching the critical point both $\lambda$ and $\Lambda$ diverge, and at the critical point power laws $P_{AA}(r) \sim 1/r^\alpha$ and $P_{AB}(r) \sim 1/r^\beta$ 
obtain, with $\alpha \approx \beta = 0.3 \pm 0.1$. In the DQC theory it is predicted that the divergence of $\Lambda$ should be faster than the correlation length 
$\xi$; $\Lambda \propto \xi^{1+k}$, with $k > 0$ and less than the exponent governing the rate of divergence of the cross-over length-scale of the emergent U($1$) 
symmetry (which should be the largest length-scale) \cite{Senthil2}. We extract $\xi$ from spin-spin correlations.  
The length-scales $\xi$ and $\Lambda$ are graphed versus the coupling ratio in Fig.~\ref{fig4}. Using lattice sizes $L$ up to $128$, we can reliably extract 
$\Lambda$ and $\xi$ when they are roughly less than $10$---beyond which the size-dependence becomes significant and extrapolating to infinite size is difficult. 
Although we can therefore not reach far into the asymptotic scaling regime, the dependence on $g-g_c$ is still consistent with the expected power-law divergence, 
if we allow a constant correction, i.e., fitting to the forms $\Lambda = a+b(g-g_c)^{-\mu}$ and $\xi = a+b(g-g_c)^{-\nu}$. We then find $\mu = 0.7(1)$ and 
$\nu = 0.8(1)$. The correlation-length exponent $\nu = 0.59(2)$ was obtained in Ref.~\cite{Sandvik2} based on finite-size scaling collapse for larger systems in 
the close neighborhood of the critical point. The results based on Fig.~\ref{fig4} have large error bars and may also be affected by further non-asymptotic 
corrections. Regardless of the precise values of $\mu$ and $\nu$, it is clear that $k=\mu/\nu-1$ is very close to $0$. This is consistent with the value 
$0.20(5)$ obtained in Ref.~\cite{Lou1} for the exponent governing the U($1$) to Z$_4$ cross-over. The exponent describing the divergence of $\lambda$ in
Fig.~\ref{fig4} is $1.1(3)$; within error bars equal to $\Lambda$.

\begin{figure}
\centerline{\includegraphics[width=7.3cm, clip]{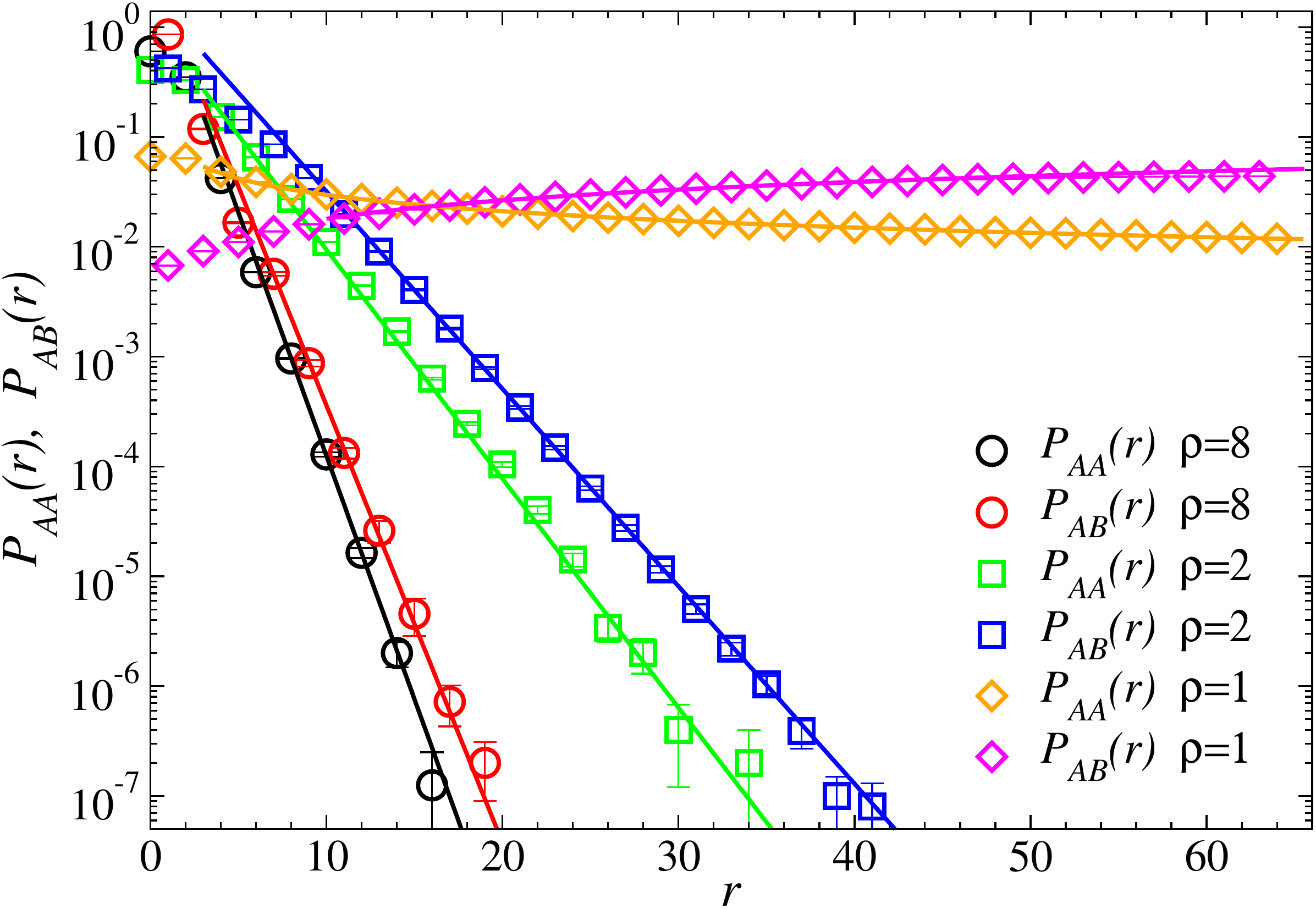}}
\vskip-2mm
\caption{(Color online) Spinon distributions in the $J_1$-$J_2$-$Q_3$ chain with $Q_3/J_1=0.1645$ (the critical point when $J_1=J_2$) for
three ratios $\rho=J_2/J_1$. The straight lines for $\rho>1$ are exponential fits and the curves for $\rho=1$ are power-law fits.}
\label{fig5}
\vskip-3mm
\end{figure}

{\it 1D deconfinement}---In the 1D VBS phase, without enforced dimerization, $J_1=J_2$ in Eq.~(\ref{j1j2}), the spinons are small and 
deconfined, as shown in Ref.~\cite{Tang1}. By turning on a symmetry-breaking dimerization, $\rho=J_2/J_1>1$, one can tune the confinement length from 
arbitrarily large to arbitrarily small \cite{Doretto,Tang3}. Here, to compare with the 2D model approaching its critical point,
we instead show results for $g=Q_3/J_1$ fixed at the critical value $g_c=0.1645$ when $J_1=J_2$ (where spinons are deconfined). Keeping $g=g_c$ and turning on the static dimerization, $\rho > 1$, 
we observe in Fig.~\ref{fig5} that $\Lambda \approx \lambda$ ($\Lambda$ being slightly larger), with both lengths diverging as $\rho \to 1$. This is similar
to the behavior observed in the 2D model (apart from the lack of spinon shrinkage). Thus, {\it the fact that the spinon size and the confinement length are 
both divergent does not invalidate the deconfinement picture}. At the critical point, the 1D ``spinon shape'' distribution $P_{AA}(r)$ also 
decays as a power-law, while the pair distribution $P_{AB}(r)$ is peaked at the longest distance, reflecting marginal (critical) spinons subject to weak repulsive 
interactions \cite{Tang1}. The main difference in the 2D model is that the effective spinon-spinon interactions are attractive, not only 
in the VBS phase but also at criticality.

{\it Conclusions and Discussion}---We showed that a spinon in the 2D $J$-$Q_3$ model shrinks when a bound-state (triplon) is formed. 
We should stress here that the reason why the vortices do not annihilate is that we restrict the system to the $S=1$ sector, while spinon 
annihilation would bring it back to the $S=0$ ground state. Both the spinon size and the confinement length diverge as the critical VBS--N\'eel 
point is approached, and at the critical point the distribution functions decay as power laws. This is also necessary for continuity, because in 
the N\'eel state both distributions become flat (as we discussed for 1D systems in Ref.~\cite{Tang1} and have also verified in 2D), when the 
spinons completely loose their identity as individual objects. 

Our scenario deviates from a simple picture of the near-critical triplon being a large object formed by two small particles. The question then is: Are the spinons 
nevertheless deconfined, independently propagating excitations? We showed with a known example that in 1D that is possible. The most plausible scenario in 2D is 
that the behaviors found here are due to weak attractive interactions between a spinon and an anti-spinon---as would in fact be expected based on 
the DQC theory, where interactions are mediated by a gauge field (but the sign
of the effective spinon-spinon interactions is not immediately clear). Note that we have only studied the lowest $S=1$ state, and in higher states 
the kinetic energy should overcome the weak attraction, leading to essentially independent spinons, as found when comparing the critical $J$-$Q$ model with a 
gas of free bosonic $S=1/2$ excitations \cite{Sandvik3}. 

In the simple RVB spin liquid, we also found weak attractive interactions, but in this case the spinons are small. The differences between the RVB liquid
and the critical $J$-$Q_3$ model should be due to the fact that the spin correlations decay exponentially in the former (and there is a spin gap), while 
they have a power-law decay the latter. Both models exhibit algebraic dimer correlations, which should be responsible for the residual spinon 
interactions. In the 1D case, full deconfinement can be seen thanks to repulsive interactions, although the critical spinons themselves are 
not small particles (as in the 2D $J$-$Q_3$ model), being instead marginally localizable objects described by power-law overlaps. Still, it is rigorously 
known that these marginal particles do propagate as individual $S=1/2$ degrees of freedom \cite{Faddeev}. 

The excitations of collective quantum states only depend on the nature of the ground state, regardless of microscopic details. We therefore expect 
our results to be generic to 2D columnar VBS states and N\'eel--VBS critical points. Our results suggest that near-critical and critical VBS are 
close 2D analogues to the 1D critical spin chains, with the differences essentially due to the different topological aspects of the spinons; 
vortices versus kink solitons.

{\it Acknowledgments---}We would like to thank C. Batista, A. Chaves, P. Fendley, and M. M\"uller for discussions. 
This research was supported by the NSF under Grant No.~DMR-1104708.

\null


\vskip-5mm
\begin{thebibliography}{00}

\bibitem{Munzinger}
P. Braun-Munzinger and J. Wambach, Rev. Mod. Phys., {\bf 81}, 1031 (2009).

\bibitem{Read}
N. Read and S. Sachdev, Phys. Rev. B {\bf 42}, 4568 (1990). 

\bibitem{Shastry}
B. S. Shastry and B. Sutherland, Phys. Rev. Lett. {\bf 47}, 964 (1981). 

\bibitem{Su}
W. P. Su, J. R. Schrieffer, and A. J. Heeger, Phys. Rev. Lett. {\bf 42}, 1698 (1979).

\bibitem{Tennant}
D. A. Tennant, T. G. Perring, R. A. Cowley, and S. E. Nagler, Phys. Rev. Lett. {\bf 70}, 4003 (1993).

\bibitem{Lake}
B. Lake, D. A. Tennant, C. D. Frost, and S. E. Nagler, Nature Mater. {\bf 4}, 329 (2005).

\bibitem{Pereira}
R. G. Pereira, J. Sirker, J.-S. Caux, R. Hagemans, J. M. Maillet, S. R. White, and I. Affleck
Phys. Rev. Lett. {\bf 96}, 257202 (2006).

\bibitem{Faddeev}
L. D. Faddeev and L. A. Takhtajan, Phys. Lett. {\bf 85}A, 375 (1981).

\bibitem{Sandvik4}
A. W. Sandvik and R. R. P. Singh, Phys. Rev. Lett. {\bf 86}, 528 (2001).

\bibitem{Zhou}
Y. Zhou and P. A. Lee, Phys. Rev. Lett. {\bf 106}, 056402 (2011).

\bibitem{Poilblanc}
D. Poilblanc, A. Laeuchli, M. Mambrini, and F. Mila, Phys. Rev. B {\bf 73}, 100403(R) (2006).

\bibitem{Doretto}
R. L. Doretto and M. Vojta, Phys. Rev. B {\bf 80}, 024411 (2009). 

\bibitem{Yan}
S. Yan, D. A. Huse, and S. R. White, Science {\bf 332}, 1173 (2011).

\bibitem{Tang1}
Y. Tang and A. W. Sandvik, Phys. Rev. Lett. {\bf 107}, 157201 (2011).

\bibitem{Senthil1}
T. Senthil, A. Vishwanath, L. Balents, S. Sachdev,  and M. P. A. Fisher, Science {\bf 303}, 1490 (2004).

\bibitem{Senthil2}
T. Senthil, L. Balents, S. Sachdev, A. Vishwanath, and M. P. A. Fisher,
Phys. Rev. B {\bf 70}, 144407 (2005).

\bibitem{Levin}
M. Levin and T. Senthil, Phys. Rev. B {\bf 70}, 220403 (2004).

\bibitem{Sachdev}
N. Read and S. Sachdev, Phys. Rev. Lett. {\bf 66}, 1773 (1991).

\bibitem{Sandvik1}
A. W. Sandvik, Phys. Rev. Lett. {\bf 98}, 227202 (2007).

\bibitem{Kaul}
R. Kaul, R. G. Melko, Phys. Rev. B. {\bf 78}, 014417 (2008).

\bibitem{Lou1} 
J. Lou, A. W. Sandvik, and N. Kawashima, Phys. Rev. B. {\bf 80}, 180414(R) (2009).

\bibitem{Sandvik2} 
A. W. Sandvik, Phys. Rev. Lett. {\bf 104}, 177201 (2010).

\bibitem{Banerjee} 
A. Banerjee, K. Damle, and F. Alet, Phys. Rev. B {\bf 82}, 155139 (2010).

\bibitem{Sandvik3}
A. W. Sandvik, V. N. Kotov, and O. P. Sushkov, Phys. Rev. Lett. {\bf 106}, 207203 (2011).

\bibitem{Anderson1}
P. W. Anderson, Mater. Res. Bull. {\bf 8}, 153 (1973).

\bibitem{Sutherland}
B. Sutherland, Phys. Rev. B {\bf 37}, 3786 (1988).

\bibitem{Wang}
L. Wang and A. W. Sandvik, Phys. Rev. B {\bf 81}, 054417 (2010).

\bibitem{Banerjee2}
A. Banerjee and Kedar Damle, J. Stat. Mech. ({\bf 2010}) P08017.

\bibitem{Liang}
S. Liang, B. Doucot, and P. W. Anderson, Phys. Rev. Lett. {\bf 61}, 365 (1988).

\bibitem{Lin}
Y.-C. Lin, Y. Tang, J. Lou, and A. W. Sandvik, Phys. Rev. B {\bf 86}, 144405 (2012).

\bibitem{Sandvik5}
A. W. Sandvik and H. G. Evertz, Phys. Rev. B {\bf 82}, 024407 (2010).

\bibitem{Tang2}
Y. Tang, A. W. Sandvik, and C. L. Henley, Phys. Rev. B {\bf 84}, 174427 (2011).

\bibitem{Alet}
A. F. Albuquerque and F. Alet, Phys. Rev. B {\bf 82}, 180408 (2010).

\bibitem{Sanyal}
S. Sanyal, A. Banerjee, and K. Damle, Phys. Rev. B {\bf 84}, 235129.

\bibitem{Anderson2}
P. W. Anderson, Science, {\bf 235}, 1196 (1987).

\bibitem{Cano}
J. Cano and P. Fendley, Phys. Rev. Lett. {\bf 105}, 067205 (2010) .

\bibitem{Murthy} 
G. Murthy and S. Sachdev, Nucl. Phys. B. {\bf 344}, 557 (1990).

\bibitem{Motrunich} 
O. I. Motrunich and A. Vishwanath, Phys. Rev. B 70, 075104 (2004).

\bibitem{Jiang} 
F.-J. Jiang, M. Nyfeler, S. Chandrasekharan, and U.-J. Wiese, J. Stat. Mech. (2008) P02009.

\bibitem{Kuklov1}
A. B. Kuklov, N. V. Prokof’ev, B. V. Svistunov, and M. Troyer, Ann. Phys. (N.Y.) {\bf 321}, 1602 (2006).

\bibitem{Kuklov2}
A. B. Kuklov, M. Matsumoto, N. V. Prokof\'ev, B. V. Svistunov, and M. Troyer, Phys. Rev. Lett. {\bf 101}, 050405 (2008).
 
\bibitem{Chaves1}
A. Chaves, F. M. Peeters, G. A. Farias, and M. V. Milo\v{s}evie\'c , Phys. Rev. B. {\bf 83}, 054516 (2011).

\bibitem{Chaves2}
A, Chaves, private communication.

\bibitem{Babaev}
E. Babaev, J. J\"aykk\"a and M. Speight, Phys. Rev. Lett. {\bf 103}, 237002 (2009).

\bibitem{Tang3}
Y. Tang and A. W. Sandvik (in preparation).

\end{thebibliography}
\end{document}